\documentstyle[12pt]{article}

\begin{document}
\makeatletter
\def\gsim{\compoundrel >\over\sim}
\def\lsim{\compoundrel <\over\sim}
\def\compoundrel#1\over#2{\mathpalette\compoundreL{{#1}\over{#2}}}
\def\compoundreL#1#2{\compoundREL#1#2}
\def\compoundREL#1#2\over#3{\mathrel
    {\vcenter{\hbox{$\m@th\buildrel{#1#2}\over{#1#3}$}}}}
\makeatother
\newcommand{\mathbold}[1]{\mbox{\boldmath$\bf#1$}}
\begin{titlepage}
\vspace*{2cm}
\begin{center}
{\bf On the width of molecular states}\\[0.5cm]
M.Shmatikov\\[0.5cm]
{\it Russian Research Center ``Kurchatov Institute'', 123182 Moscow,
Russia}\\[1cm]
{\bf Abstract}\\[0.25cm]
\end{center}
Width of molecular-type $Q\bar{Q}q\bar{q}$ hadronic states is discussed.
It depends on the quantum numbers of the state and on the details of the
long-range mechanism ensuring the bindedness of a molecule. Specific
cases when the width is large are considered in detail.
\end{titlepage}
Existence of multiquark exotic $QQ\bar{q}\bar{q}$ states attracts
interest for a long time since at present there are no evidence
for any other states contaning more that 3 quarks which are stable
with respect to strong decays. The possibility that such
states could be bound for large enough $m_Q/m_q$ mass ratio
was shown in \cite{Ader,Lipkin} to emerge as a consequence
of flavor independence of chromoelectric forces. The interest to
the problem was rekindled recently by the studies of
of $QQ\bar{q}\bar{q}$ states in the framework of
chiral perturbation theory \cite{Manohar}.
Qualitative conclusions of \cite{Manohar} can be formulated as follows:
1)~in the infinite mass limit ($m_q\rightarrow\infty$) stable
$QQ\bar{q}\bar{q}$ meson exists; 2)~states containing $Q=c,b$
quarks may be weakly bound by the long-range one-pion forces
which, moreover, are calculable in the chiral perturbation theory.
The same conclusion as to existence of a bound state is valid also
for a meson-antimeson $\bar{Q}qQ\bar{q}$ molecule.

It should be stressed that the infinite mass limit may be a poor
approximation even for heavy $Q =c,b$ quarks. Then corrections
$\simeq 1/m_Q$ should be taken into account, the most important one
being related to mass difference $\Delta m_Q$ between vector and
pseudoscalar heavy mesons.

Investigation of molecular states consisting of heavy mesons with
the account of finite heavy-quark corrections was carried out in
\cite{T1,T2}. One-pion exchange (or, more precisely, its
tensor component) was considered as a driving force for heavy-meson
interaction. The effective coupling constant of the pion and the
heavy meson $g$ was assumed to be universal for all heavy mesons
due to pion interaction with the isovector current of a
light quark only:
\begin{equation}
{\cal L}_{int} = \frac{g}{f_{\pi}}\,\bar{q}(x)\gamma^{\mu}\gamma_5
\vec{\tau}\,q(x)\partial_{\mu}\vec{\pi}(x)\,,
\end{equation}
where $f_{\pi}$ is the $\pi$-meson decay constant
($f_{\pi}\approx 132$~MeV).

A number of deuteron-like loosely bound states was found in the
charm and beauty sectors, the binding energy in the latter case
amounting to about 50~MeV. In the channels with exotic flavor
quantum numbers ($QQ\bar{q}\bar{q}$ states) pion exchange
proves to be either too weak or even repulsive. The $H^*\bar{H}^*$
states (where $H^*$ is the generic notation for a vector $D^*$ or $B^*$
meson) are expected to decay into a pair of pseudoscalar $H\bar{H}$
mesons and it was conjectured in \cite{T2} that the corresponding
width is about tens of MeV.

In the present paper we investigate decay mechanisms of molecular-type
heavy mesons with hidden flavor. First, such states can annihilate
into ordinary mesons. One can get readily estimates for the annihilation
width. Indeed, the radius of the annihilation ($\simeq 1/m_Q$) is much
smaller than that of the bound state which can be assumed safely
to be $\simeq 1/\Lambda$, where $\Lambda$ is the QCD parameter. Then
applying the familiar formula \cite{Deser} we arrive immediately at
\begin{equation}
\Gamma = (v\sigma_{ann})\vert\psi_{\pi}(0)\vert^2\,,
\label{gmm}
\end{equation}
where $v$ is the velocity of mesons bound in a molecular state,
$\sigma_{ann}$ is the annihilation cross section and $\psi_{\pi}$ is
the wave function of the state generated by the one-pion-exchange
force. For a loosely bound state the $\psi_{\pi}$ wave function can
be approximated as follows
\begin{equation}
\psi_{\pi}(r) = \sqrt{\frac{\kappa}{2\pi}}\,\frac{\exp(-\kappa r)}{r}
\label{wf}
\end{equation}
where $\kappa$ is the bound-state momentum related to the binding
energy $\varepsilon$: \mbox{$\kappa = \sqrt{m_Q\,\varepsilon}$}.
The value of the $(v\sigma_{ann})$ can be extracted from the width
of the pseudoscalar $Q\bar{Q}$ meson. For the positron-like system
\cite{LL}
\begin{equation}
(v\sigma_{ann}) \approx \frac{32\pi \alpha^2_S}{12 m^2_Q}
\label{ann}
\end{equation}
Combining (\ref{wf}) and (\ref{ann}) we get
\begin{equation}
\frac{\Gamma}{\varepsilon} = \sqrt{\frac{m_Q}{\varepsilon}}\cdot
\frac{4}{3}\,\frac{\alpha^2_S}{(m_Q a)^2}
\label{rat}
\end{equation}
where $a$ is the radius of the force. Taking the latter
$a \approx 1/2 m_{\pi}$ \cite{Manohar}-\cite{T2} we rewrite
(\ref{rat}) in the form
\begin{equation}
\frac{\Gamma}{\varepsilon} = \sqrt{\frac{m_Q}{\varepsilon}}\,
\frac{4}{3}\left(\alpha_S\,\frac{2 m_{\pi}}{m_Q}\right)^2
\end{equation}
yielding for the characteristic value of the binding energy
$\varepsilon\approx 10$~MeV in the
$D^*\bar{D}^*$ state
\begin{equation}
\Gamma/\varepsilon \simeq 0.1
\label{ge}
\end{equation}
For the $B^*\bar{B}^*$ state and/or nonzero orbital momenta
this ratio will be even smaller.

Finite mass of a heavy quark makes vector $H^*$ mesons heavier
than their pseudoscalar conterparts $H$. Nevertheless, for the
apparent reason decay of $H^*\bar{H}$ and  $H^*\bar{H}^*$
states with abnormal spin-parity quantum numbers
into the $H\bar{H}$ pair is forbidden. These molecules will
decay by annihilating into ordinary mesons with the width estimate
(\ref{ge}) being applicable. At the same time bound states
of heavy vector mesons $H^*\bar{H}^*$ with normal spin-parity
quantum numbers can decay into a pair of pseudoscalar mesons:
\mbox{$H^*\bar{H}^*\rightarrow H\bar{H}$}, and because of this
decay channel their width may be large.

Let us consider the $H^*\bar{H}^*$ bound state with the quantum
number \mbox{$J^{\pi} = 0^+$}, involving two partial waves ${}^1S_0$ and
${}^5D_0$. Besides the $H^*\bar{H}^*$ state can couple to the
$H\bar{H}$ pair with the $J^{\pi} = 0^+$ quantum numbers. Following
\cite{T1,T2} we bound our consideration by the most long
range part of forces operating between mesons, i.e. one-pion exchange.
Chiral invariance (at least approximately) determines coupling of
the $\pi$-meson to a heavy meson. Note that for parity reasons
the $H\bar{H}\pi$ vertex vanishes. One-pion exchange proves to be strong
enough to bind a pair of vector mesons, the mass of the bound state
$M$ being \mbox{$2m_H < M < 2m_{H^*}$} \cite{T1,T2}. Just the
same mechanism (one-pion exchange) couples the bound state to the
noninteracting $H\bar{H}$ pair. Thus we arrive at a system of 3 coupled
channels
\begin{equation}
H\bar{H}({}^1S_0) \leftrightarrow H^*\bar{H}^*({}^1S_0) \leftrightarrow
H^*\bar{H}^*({}^5D_0)
\label{ch}
\end{equation}
Forbidding complexity of the problem necessitates making some
simplifications. Having in mind the qualitative character of our
results we neglect the contribution of the last channel in (\ref{ch}),
since it provides some additional attraction only without changing
the general character of interaction in the considered system.
Then the potential matrix in the considered system reads
\begin{equation}
V = \left(
\begin{array}{cc}
0 & V_{12}\\
V_{21} & V_{22}\\
\end{array}
\right)
\label{pot}
\end{equation}
with $H\bar{H}$ and $H^*\bar{H}^*$ channels carrying numbers 1 and 2
respectively. Each component of the potential matrix can be represented
in terms of one and the same function describing one-pion exchange
\begin{equation}
V = -V_0\,(\vec{\tau}_1\cdot\vec{\tau}_2)\,
\left[
\begin{array}{cc}
0 & -\sqrt{3}\\
-\sqrt{3} & -2\\
\end{array}
\right]\,
\cdot\frac{\exp(-m_{\pi}r)}{m_{\pi}r}\,,
\label{un}
\end{equation}
where $\vec{\tau}$ is the isospin Pauli matrix and $V_0$ is the
``universal'' coupling strength constant \cite{T1,T2}:
\begin{equation}
V_0 = \frac{m^3_{\pi}}{12\pi}\,\frac{g^2}{f^2_{\pi}}\,.
\end{equation}
For the $g$ value as extracted from the $D^*\rightarrow D+\pi$ decay
($g\approx 0.6$) this constant equals $V_0\approx 1.3$~MeV.
Note that one-pion exchange ensures attraction in the
considered system in the isoscalar channel only (additional
minus sign appears in the meson-antimeson case).

To take into account approximate value of coupling constants and
investigate behavior of observables with their variation we
modify interaction potential (\ref{pot}), (\ref{un}) as follows
\begin{equation}
\begin{array}{ll}
{}& V_{12}(r) \rightarrow V_{12}(\lambda',r) = \lambda'\cdot V_{12}(r)\\
{}& V_{21}(r) \rightarrow V_{21}(\lambda',r) = \lambda'\cdot V_{21}(r)\\
{}& V_{22}(r) \rightarrow V_{22}(\lambda,r) = \lambda\cdot V_{12}(r)\\
\end{array}
\label{mod}
\end{equation}
where $\lambda$ and $\lambda'$ are some numerical constants. The
former constant controls the strength of interaction in the
first ($H^*\bar{H}^*$) channel and the latter the strength of its
coupling to the ``annihilation'' ($H\bar{H}$) channel. The
``reference'' values of the potential strengths correspond then to
$\lambda, \lambda' = 1$.

Interaction potentials $V_{22}$ and $V_{12}$ have comparable
strength and equal range precluding application of perturbation
theory approximations. Using the modified potential (\ref{pot})
and (\ref{mod}) we solve
numerically the system of coupled Schr\"odinger equation and investigate
behavior of the scattering length ${\cal A}$ in the second ($H^*\bar{H}^*$)
channel
as the function of $\lambda$ and $\lambda'$. In the case of uncoupled
``annihilation'' ($H\bar{H}$) channel (corresponding to $\lambda' = 0$)
emergence of a (loosely) bound state is signalled by
${\cal A} \rightarrow\infty$ at some critical value of the strength
coupling constant $\lambda_{cr}$. Coupling to the $H\bar{H}$ channel
drastically
changes behavior of ${\cal A}(\lambda)$. First, now molecule components
spend some time in the $H\bar{H}$ state where interaction is absent.
It implies that the average attraction in the considered system
weakens. Then formation of a bound state requires corresponding
increase of attraction in the $H^*\bar{H}^*$ state: domain of
positive ${\cal A}$'s, corresponding to bound states, shifts
to the right along the \mbox{$\lambda$-axis}.
For given value of $\lambda$ a bound state may even become
an (unbound) virtual state. Besides, the ${\cal A}(\lambda)$
curve remains continuous for all values of $\lambda$.
Second, the scattering length ${\cal A}$ acquires imaginary part.
Its magnitude and behavior also depend on $\lambda$.
The imaginary part Im~${\cal A}$ achieves its extremum in the
vicinity of the
$\lambda$ value where a bound state emerges. Dependence of both
real and imaginary parts of the ${\cal A}$ scattering length is
illustrated in fig.1 for the realistic case $\lambda' = 1$.

Inspection of curves in fig.1 shows that the
$\lambda$-behavior of the scattering length components
reproduces the pattern of a resonance system with friction:
absorption reaches its maximum just at the point where the
real part of the resonance curve vanishes \cite{Landau}.
With weakening of ``annihilation'' strength ($\lambda'$ diminishing)
the Im~${\cal A}$ peak becomes more narrow and high.
Hence the \mbox{$\rho = {\rm Im}~{\cal A}/{\rm Re}~{\cal A}$ } ratio
may be both small
and comparatively large ($\approx 1$) depending on specific
values of $\lambda$ and $\lambda'$. It should be stressed that
rapid variation of the $\rho$ ratio occurs in the region of
``realistic'' values ($\lambda,\lambda' = 1$) and the extension
of this region corresponds roughly to the accuracy with which
the interaction potentials and hence $\lambda,\lambda'$ strength
coupling constants are known. These results are quite general
and hold valid for any other (normal) spin-parity quantum
numbers of the $H^*\bar{H}^*$--$H\bar{H}$ molecular states.

Resuming, we have considered the widths of molecular states consisting
of heavy mesons. States involving vector and pseudoscalar
($H^*\bar{H}$) and vector mesons ($H^*\bar{H}^*$) with the
abnormal spin-parity are expected to be very narrow decaying
into ordinary mesons with the width given by (\ref{ge}).
Vector-meson ($H^*\bar{H}^*$) molecular states with normal
spin-parity quantum numbers will decay into the pair of
pseudoscalar mesons ($H^*\bar{H}^*\rightarrow H\bar{H}$).
The qualitative conclusion is that the width of such
states may vary in a wide range and, in particular,
they can be rather broad and hence unobservable as resonances.
The width of a resonance depends upon details of interaction
mechanism and cannot be predicted with the state-of-the-art
accuracy of coupling constants. We have considered the case
when the mass of a molecular state exceeds the double mass
of a heavy pseudoscalar meson ($M > 2m_H$). Provided the
binding energy is larger than the double mass difference
between the vector and the pseudoscalar heavy meson
($M < 2m_H$), the $H^*\bar{H}^*\rightarrow H\bar{H}$ decay
channel will be closed and these molecular states will be stable
with respect
to strong decay. It should be noted, however, that for such strong binding
characteristic momentum of the bound
state is $\kappa_H\gsim\sqrt{m_H\cdot 2\Delta m_H}\approx 700$~MeV
and the corresponding characteristic distances
$r_H\approx 1/\kappa_H\approx 0.3$~Fm. These values indicate that
interacting heavy mesons are located well within the meson's quark-core
domain and the calculations based on the analysis of long-range forces
only may prove to be unreliable.

\newpage
\begin{center}
Figure caption\\[0.5cm]
\end{center}
Fig.1. Dependence of the scattering length ${\cal A}$ [Fm] in the
$H^*\bar{H}^*$ channel upon the relative strength constant $\lambda$.
Solid curves represent behavior of the real and imaginary parts
of ${\cal A}$ for $\lambda'=1$ (full-fledged coupling to the
``annihilation'' $H\bar{H}$ channel), while dotted curve is for the
real part of ${\cal A}$ for $\lambda'=0$ (uncoupled ``annihilation''
channel: imaginary part of ${\cal A}$ vanishes). ``Reference'' values
of relative strength coupling constants are $\lambda,\lambda'=1$.
\end{document}